\documentstyle[amsbsy,amstex,aps,epsf,multicol]{revtex}

\def\eqref#1{Eq.~(\ref{#1})}

\def\phi{\varphi}

\def\({\left(}
\def\){\right)}
\def\[{\left[}
\def\]{\right]}
\def\<{\left\langle}
\def\>{\right\rangle}
\def\<{\left\langle}
\def\>{\right\rangle}

\def\bea{\begin{eqnarray}}
\def\eea{\end{eqnarray}}

\makeatletter

\title{Supersonic turbulence and  structure of interstellar 
molecular clouds}
\author{Stanislav Boldyrev\\
{\em Institute for Theoretical Physics,
Santa Barbara, California 93106}, {\sf boldyrev@itp.ucsb.edu} }
\author{{\AA}ke Nordlund\\
{\em  Copenhagen Astronomical Observatory and Theoretical
Astrophysics Center, DK-2100 Copenhagen, Denmark}, {\sf aake@astro.ku.dk}}
\author{Paolo Padoan\\
{\em Jet Propulsion Laboratory, 4800 Oak Grove Drive, 
MS 169-506,\\
California 
Institute of Technology, Pasadena, CA 91109-8099,
} 
{\sf padoan@jpl.nasa.gov}}

\date{March 25, 2002}

\begin{document}

\input psfig.sty


\maketitle

\begin{abstract}
\vskip5mm
The interstellar 
medium (ISM) provides a unique laboratory for highly supersonic, 
driven  hydrodynamics turbulence. We present a theory of such 
turbulence,  confirm it by numerical simulations, and use the 
results to explain 
observational properties of interstellar molecular clouds, the 
regions 
where stars are born.


\end{abstract}

\begin{multicols}{2}

{\bf 1.} {\em Introduction.} Stars are formed as a result 
of gravitation (Jeans) collapse of dense clumps 
in interstellar molecular clouds. The structure of 
such clouds in a large interval of scales (from about $100pc$ to
$0.01pc$) lacks any
characteristic length, and 
can be understood as arising from supersonic hydrodynamic motions 
sustained on 
large scales by supernovae explosions~\cite{Elmegreen1,avillez,Korpi}. 
A fluid motion with characteristic 
large-scale relative velocities of order~1-10 km/sec compresses 
rapidly 
radiating and therefore relatively cold molecular gas ($T\sim 10K$) up 
to very high densities (above~$10^4 cm^{-3}$). Instabilities 
of shock fronts create 
a hierarchy of gas clumps with broad mass, size, and velocity 
distribution 
controlled by the Mach number ($M=v/c$) and by the Alfv\'enic Mach 
number~($M_a=v/v_a$), where $v_a$~is the Alfv\'en 
velocity,~$v_a=B/\sqrt{4\pi\rho}$, and $c$~is the sound speed. 
Depending on parameters of  
clouds, the Mach number can be about 30 on the largest scales, 
and the Alfv\'enic Mach number can exceed 1 as
well~\cite{Larson1,Larson,Falgarone,Falgarone1,Myers,Padoan,Ossenkopf,Ostriker}.  
A systematic study of scaling properties of 
supersonic turbulence and of the structure of molecular clouds 
was initiated by the work of Larson~\cite{Larson1,Larson}, but 
despite the 
large number of observational and numerical investigations
that appeared 
for the last 20 years, the theoretical understanding of the turbulence 
has been rather poor.

In the present paper we provide a new analytical model for 
a supersonic 
turbulent cascade  
and  test the model against observations and 
  numerical simulations. We  
find a very good 
agreement within the error bar uncertainty. The analytical 
approach is suggested by the following two numerical results. First 
of all, we establish that for large Mach numbers ($M>2$) 
the velocity field in the inertial interval is mostly divergence-free
and shear-dominated, 
with the intensity of its potential component being only about 10\% 
of the intensity of its solenoidal
part. The second finding  
is that the most intense dissipative structures of the turbulence 
are two-dimensional sheets or shocks as oppose to the incompressible 
case where most of energy is dissipated in
filaments, see also~\cite{Ostriker,Porter1}. Using these two 
ingredients in the framework of the so-called She-L\'ev\^eque 
model 
of turbulence we calculate two-point correlators of velocity and 
density fields. This allows us to construct the 
multifractal distributions of velocity and density fields, 
that statistically describe the structure of
a turbulent molecular cloud. In the present paper 
we present the analytic derivation of the model and 
summarize the most important numerical results. The detailed 
numerical and observational analysis will appear
elsewhere~\cite{BNP,Padoan1}.

In the following section we construct the multifractal distribution 
of the velocity field, and 
numerically check the first~10 velocity-difference structure 
functions [the definition is given below]. In section~{\bf 3} we 
proceed with the density distribution, derive 
a general formula for two-point density correlators 
and compare the result 
with the numerical simulations. Conclusions, applications, 
and future research are outlined in section~4.

{\bf 2.} {\em Multifractal model of supersonic turbulence}. 
In 1994 She and L\'ev\^eque suggested a model that turned out to be 
very successful in explaining the experimental findings for 
incompressible turbulence~\cite{She_Leveque1}. 
The model represents a turbulent cascade as an infinitely-divisible 
log-Poisson process~\cite{Dubrulle,She_Leveque2,Novikov}, and 
has three input parameters. The first two of 
them are the so-called {\em naive}, i.e., non-intermittent, 
scaling exponents of velocity fluctuation and of 
the eddy turnover time. The 
non-intermittent velocity of the eddy of size~$l$ scales 
as~$v_l\sim l^{\Theta}$, and the ``eddy turnover" time scales 
as~$t_l\sim l^{\Delta}$. The third parameter is the dimension of the 
most intense dissipative structure,~$D$. In the incompressible case, 
the first two parameters take their Kolmogorov 
values, $\Delta=2/3$, $\Theta=1/3$, and the third parameter is~$D=1$ 
since the dissipation mostly occurs in elongated filaments. 

The model 
predicts the so-called structure functions of the velocity field that 
are defined as follows
\begin{eqnarray}
S_n(l)=\langle |v(x+l)-v(x)| \rangle \sim l^{\zeta(n)}.
\label{structure_functions}
\end{eqnarray}
The velocity components in this formula can be either parallel or 
perpendicular to vector~$l$. In the former case the structure 
function is called longitudinal, in the latter case transversal. 
There is experimental evidence that both scale in the same
way~\cite{Benzi2,Camussi,Frisch}, 
therefore, we will not distinguish between them as far as the
scaling is concerned.  
The She-L\'ev\^eque 
formula gives the following expression for the scaling exponents of 
the structure functions,
\begin{eqnarray}
\zeta(n)=\Theta (1-\Delta)n +(3-D)(1-\Sigma^{\Theta n}),
\label{she_leveque}
\end{eqnarray}
where~$\Sigma=1-\Delta/(3-D)$~\cite{She_Leveque1}. 
This expression has been 
experimentally checked to work for structure functions up to the 
10th order, within  an accuracy of a few 
percent~\cite{Benzi2,She_Leveque1}. 

To address the {\em supersonic} case, we note that in 
the inertial interval the turbulence is still mostly divergence-free
(Fig.~(\ref{energies})).
The effect of vorticity generation in 
the random 3D~flow is 
analogous to the effect of magnetic dynamo, since magnetic field 
and vorticity obey the same dynamic equation. We therefore 
leave the Kolmogorov parameters $\Theta$ and $\Delta$ unchanged. 
{
\columnwidth=3.2in
\begin{figure} [tbp]
\centerline{\psfig{file=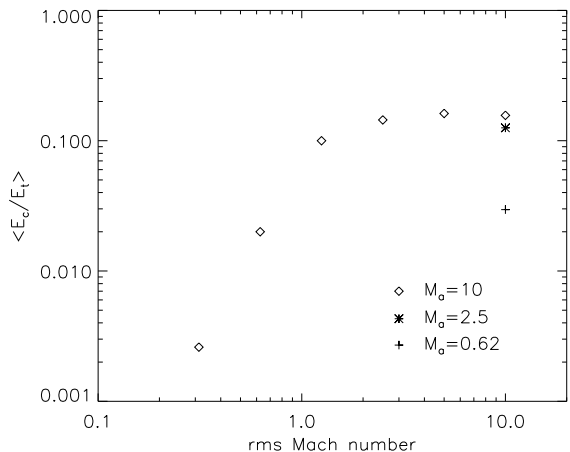,width=3.2in}}
\caption{Results from numerical simulations of randomly driven 
MHD equations with resolution~$128^3$. 
The ratios of the potential, $E_c$, to the solenoidal,$E_t$, 
part of the velocity field are 
presented for the Mach numbers up to~$M=10$ at the
largest scale. 
The initial magnetic Mach number has been chosen in the
range~$M_a=0.6,\dots,10$, 
and the numerical integration has been conducted for several turn-over 
times of the largest eddies, which was enough to 
reach the steady state.   
The isothermal equation of state was used and the large-scale driving 
force was 
Gaussian, solenoidal, and correlated at a turn-over time of the 
largest eddy.
The detailed description of the numerical set up can be found in~[4].
The scaling relations reported in the present 
paper were obtained for~$M\simeq 10$, and~$M_a \simeq 3$. 
} 
\label{energies}
\end{figure}
}
However, the most dissipative structures of a supersonic flow are 
different from the incompressible case.
Instead of filaments 
they look like shocks or two-dimensional dissipative sheets, 
therefore, $D=2$~\cite{Boldyrev,BNP}. 
With such 
input parameters, formula~(\ref{she_leveque}) is recast as
\begin{eqnarray}
\zeta(n)=n/9+1-\left(1/3 \right)^{n/3}.
\label{she_leveque1}
\end{eqnarray}
Quite remarkably, this formula works with good 
accuracy for the numerical simulations, fitting the first ten 
structure functions with an error of about~5\%, see Fig.(\ref{p10}).
{
\columnwidth=3.2in
\begin{figure} [tbp]
\centerline{\psfig{file=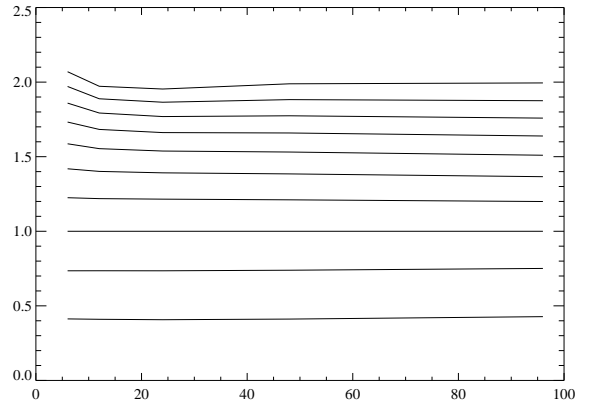,width=3.3in}}
\caption{Slopes of transversal structure functions 
computed for~$n=1,2,\dots, 10$
(correspondingly, from bottom to top), in the $250^3$ run with~$M=10$
and~$M_a=3$. 
The plot presents the ratios of the {\em differential} slopes of the
structure functions 
to the differential slope of~$S_3(l)$. These ratios,
$\zeta(n)/\zeta(3)$, exhibit excellent 
scalings, in agreement with the
Extended Self-Similarity hypothesis~[2].
They are well
described by our formula~(\ref{she_leveque1}). Note the strong 
difference of the scalings of the structure functions 
from the scalings given by the Kolmogorov model, $\zeta(n)=n/3$, and 
by the Burgers model, $\zeta(n)=1$, see~[15].}
\label{p10}
\end{figure}
}
The observational data consistently indicate steeper than 
 Kolmogorov velocity spectra, see
e.g.~\cite{Ossenkopf,Falgarone1}. 
Our model~(\ref{she_leveque1}) 
predicts~$|v_k|^2\sim
k^{-1-\zeta(2)}=k^{-1.74}$, which agrees with observations within 
error bars.
Interestingly enough, the average velocity spectrum,
originally inferred from
observations by Larson, was~$k^{-1.74}$ on scales of order~$1pc <
l<1000pc$,
and~$k^{-1.76}$ towards smaller
scales,~$0.1pc<l<100pc$~\cite{Larson,Larson1}. 

On the analytical side, the model~(\ref{she_leveque1}) 
implies  the so-called 
multi-fractal distribution of the turbulent fluctuations. Here
we  discuss the statistics of   
the velocity field,  and in the next section, apply the results 
to the density field.
To visualize the model, assume that the whole 
space (a molecular cloud or a simulation domain) contains turbulent 
structures of different (in general, fractal) dimensions. In 
the vicinity of a particular structure, the velocity difference 
scales with some particular 
exponent that we denote by~$h$, i.e., $v_l\sim l^h$. The dimension 
of this fractal structure will be denoted~$D(h)$. If we divide 
the space into small boxes of size~$l$, then the number of boxes 
covering the fractal structure of dimension~$D$ is
proportional to~$l^{-D}$, 
while the total number of boxes is proportional to~$l^{-3}$. 
The probability to 
find ourselves inside a box covering the fractal with 
dimension~$D(h)$ is 
therefore~$p_h(l)\sim l^{3-D(h)}$. To average the $n$'s moment of the 
velocity difference we just need to sum~$l^{nh} p_h(l)$, which 
is the contribution of one particular fractal structure, over 
all the fractal structures. We get
\begin{eqnarray}
S_n(l) \sim \sum\limits_h l^{nh + 3-D(h)}\sim l^{\zeta(n)}.
\label{multifractal}
\end{eqnarray}
Knowing~$D(h)$ is equivalent to knowing~$\zeta(n)$, these two 
functions are related by the Legendre transform as one immediately 
gets by assuming that~$l$ is small, and by evaluating the 
sum in~(\ref{multifractal}) by the steepest 
descent method~\cite{Frisch}. For example, knowing~$\zeta(n)$
from~(\ref{she_leveque1}), one can restore~$D(h)$; we, however, will
not need~$D(h)$ for our present purposes.

{\bf 3.} {\em Multifractal distribution of the density field}. 
First we would like to derive an important relation of  
supersonic 
turbulence. We start with the Navier-Stokes and continuity equations:
\begin{eqnarray}
& \partial_t {\bf u} + ({\bf u}\cdot \nabla) {\bf u}=
({\eta}/{\rho}) \Delta {\bf u}-c^2 \nabla \rho/\rho +{\bf f},
\label{navier-stokes}\\
& \partial_t \rho + \nabla \cdot (\rho {\bf u})=0.
\label{continuity}
\end{eqnarray}
Let us introduce the density correlator 
$R(t,x)=\langle \rho(x_1,t)\rho(x_2,t) \rangle$, 
and the density-weighted second-order velocity structure
function, 
$G^{ik}(t,x)=\langle \rho(1) \rho(2) [u(1)-u(2)]^i 
[u(1)-u(2)]^k \rangle$,  
where~${\bf x}={\bf x}_1-{\bf x}_2$. Averaging is performed over 
the random force.
Differentiating~$R(t,x)$ with respect to time, and
using~(\ref{navier-stokes}) and~(\ref{continuity}), 
one gets: 
\begin{eqnarray}
& \partial^2_t R + \nabla_i \nabla_k G^{ik} 
-c^2 \Delta R + \nabla_i \langle [f(1)-f(2)]^i\rho(1) \rho(2) 
\rangle \nonumber \\ 
&= 2 \eta \Delta \nabla_i \langle u(1)^i \rho(2)
 \rangle,   
\end{eqnarray}
where the spatial derivatives are taken with respect to~${\bf x}$. 
In the 
inertial interval, the forcing and the viscous terms are small. In the 
supersonic regime, one can also neglect the~$c^2$ term. Assuming now 
that the turbulence is in a steady state, we are left with a  
simple equation that must hold in the inertial interval, 
$\nabla_i \nabla_k G^{ik}=0$. Due to spatial isotropy 
we get:
\begin{eqnarray}
G^{ik}(x)=A\delta^{ik}+O(x^2/L^2)+\dots,
\label{restriction}
\end{eqnarray}
where~$L$ is the external force correlation length, and~$A$ is some
constant. In the inertial
interval, $x\ll L$, one gets~$G=const$. 
To obtain the density distribution 
let us make a natural   
assumption that both the density and the velocity 
fields 
have fixed scalings in the vicinity of the {\em same} turbulent  
structures. 
In other words, close to a fractal structure where the velocity 
field has 
scaling~$h$, the density field 
has some other, but also constant along the same structure, 
scaling~$\alpha (h)$, i.e.,~$\rho(l)\sim l^{\alpha(h)}$. The 
condition~(\ref{restriction}) now reads
\begin{eqnarray} 
G \sim \sum_h l^{2h + 2\alpha(h)+3-D(h)}=const.  
\label{restriction1}
\end{eqnarray}
The other restriction comes from the mass conservation law,
\begin{eqnarray}
\langle \rho \rangle \sim \sum_h l^{\alpha(h)+3-D(h)}=const.
\label{restriction2}
\end{eqnarray}
Strictly speaking, our constraint
conditions~(\ref{restriction1}) and~(\ref{restriction2}) are 
to a certain extent phenomenological. However, we found them to be 
consistent with our simulations. Moreover, the theory based on 
them predicts density correlators rather successfully, which we 
are going to demonstrate now.  Let us 
assume that the function~$\alpha(h)$ is analytic and can be 
expanded as~$\alpha(h)=a+bh+gh^2+\dots$. As a minimal model, consider 
the 
case, when~$\alpha(h)$ is a {\em linear} function, $\alpha(h)=a+bh$. 
It 
turns 
out that such a linear ansatz is consistent with both restriction 
conditions~(\ref{restriction1}) and~(\ref{restriction2}).  As follows 
from~(\ref{multifractal}), the mass conservation   
condition~(\ref{restriction2}) is satisfied 
with~$a=-\zeta(p)$ and~$b=p$, where~$p$ is 
arbitrary.
The equation for~$p$ is then derived from the
dynamic constraint~(\ref{restriction1}), which 
gives ~$\zeta(2p+2)=2\zeta(p)$. 
The solution of this equation will be denoted as~$p_0$. 
If we use our formula~(\ref{she_leveque1}), $p_0$ 
can be found exactly; 
we thus  
obtain~$b=p_0=2.28$ and ~$a=-\zeta(p_0)=-0.82$. 
The multifractal distribution for the 
density field,~$D_{\rho}(\alpha)$, is thus related to 
the multifractal 
distribution of the  velocity field,
$D_{\rho}(\alpha)=D\left[{(\alpha+\zeta(p_0))}/{p_0} \right]$. 
Fractal
and multifractal distributions of density fields 
have indeed been inferred from observations,
see, e.g.,~\cite{Elmegreen,Chappell}, and from numerical
simulations~\cite{Vazquez}.

By analogy with the velocity field, the quantities of practical 
interest are density correlators. They can be calculated with 
the aid of a formula analogous to~(\ref{multifractal}),
\begin{eqnarray}
 \langle [\rho(x+l)\rho(x)]^m \rangle 
 \sim \sum\limits_h 
l^{2m\alpha(h)+3-D(h)}\sim l^{\xi(m)}.
\label{density_structure}
\end{eqnarray}
Upon substituting the linear expression for~$\alpha(h)$ and using
formula~(\ref{multifractal}), one immediately 
gets $\xi(m)=\zeta(2 m p_0)-2m\zeta(p_0)$. This formula allows one 
to obtain the density scaling 
if the velocity structure functions are known either from theory or 
from experiment.
Since~$\zeta(n)$ is a concave 
function, $\xi(m)$ is negative for~$m>1/2$.
For $m=1,2,3$ the formula gives~$\xi(1)\simeq -0.3$,  
$\xi(2)\simeq -1.3$, and $\xi(3)\simeq -2.4$, 
values close to what is obtained in the numerics,
see Fig.~(\ref{density}). 
We give here only the first three exponents since starting from~$m=3$, 
the density exponents depend on velocity structure functions of 
order higher than 13, which cannot be reliably produced with our 
numerical resolution. 

{\bf 4.} {\em Conclusions}. We have suggested a self-consistent model 
that provides an
explanation for numerical and observational scaling laws of 
supersonic
ISM turbulence, the so-called Larson's laws. 
We would like to conclude with the following remarks:
\begin{enumerate}
\item The She-Leveque approach was also applied to 
{\em incompressible} MHD
turbulence in~\cite{Grauer,Politano,Biskamp}, where different 
scalings were 
suggested. Most successful was the approach of M\"uller
and Biskamp~\cite{Biskamp}, where 
the energy cascade was assumed to
be Kolmogorov-like, but the dissipation occurred in micro current 
sheets. 
In this case, the same formula~(\ref{she_leveque1}) gave a good 
agreement with numerical simulations for structure functions up to
order~8. 
Our results together with those by M\"uller and Biskamp 
support the ideas put forward 
in~\cite{Dubrulle} and~\cite{She_Leveque2}, that completely
different turbulent systems can belong to the same class of 
universality, i.e.
have the same velocity scaling exponents.
\item Turbulence with small pressure is usually referred to as Burgers
turbulence, the theory of which has been rapidly developing in recent
years~\cite{Polyakov,Yakhot,Boldyrev1,Weinan,Gotoh,Verma,Frisch1,Passot}.
 However, the Burgers velocity field is usually assumed to be  
potential, which is true in one and two
dimensions, but inconsistent with the 3D case due to strong
vorticity generation. However, our general 
relation,~$\rho(1)\rho(2)\sim |v(1)-v(2)|^{2p_0}/l^{2\zeta(p_0)}$, 
which is valid inside any correlation function, can be useful 
for the closure problems of Burgers turbulence.
\item Our model is consistent with available observational results,
although the error bars of observed velocity structure functions  
are too large for a precise comparison.
Moreover, only projected quantities (i.e., integrated along the line of
sight) are observationally available, 
and therefore the 3D results should be reformulated for these 
projected
fields --- this is a subject of future
work~\cite{Padoan1}.
\end{enumerate}
{
\columnwidth=3.2in
\begin{figure} [tbp]
\centerline{\psfig{file=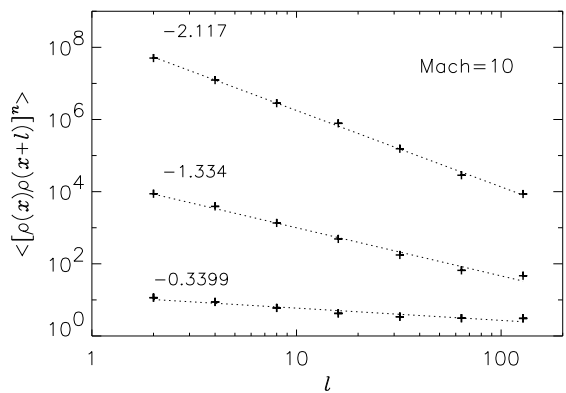,width=3.3in}}
\caption{Density correlators for~$m=1,2,3$. 
The numerically obtained slopes
are~$\xi(1)\simeq-0.3$, $\xi(2)\simeq -1.3$, and~$\xi(3)\simeq-2.1$,
close to the theoretical prediction~(\ref{density_structure}). 
The numerical simulations are the
same as in Fig.~(\ref{energies}).}
\label{density}
\end{figure}
}

\end{multicols}

\end {document}